\documentclass[a4paper,11pt]{article}

\pdfoutput=1

\newcommand{\defaultlinespread}{\linespread{0.94}}

\usepackage{geometry}
\defaultlinespread{}
\newenvironment{mtwocols}{\begin{multicols}{2}}{\end{multicols}}

\usepackage{array} 
\newcolumntype{L}[1]{>{\raggedright\let\newline\\\arraybackslash\hspace{0pt}}m{#1}}
\newcolumntype{C}[1]{>{\centering\let\newline\\\arraybackslash\hspace{0pt}}m{#1}}
\newcolumntype{R}[1]{>{\raggedleft\let\newline\\\arraybackslash\hspace{0pt}}m{#1}}

\newcolumntype{M}{>{\centering\arraybackslash$$}p{.985\linewidth}<{$$}}

\usepackage{environ}
\NewEnviron{mtvDisplayMath}{%
\vspace{4mm}
\noindent\colorbox{magenta!43!cyan!10!white}{%
\begin{tabular}{@{}M@{}} \\[-24mm]
\BODY \\[-1mm]
\end{tabular}}
\vspace{1mm}
}

\usepackage{amsmath}
\usepackage{amsfonts} % \Box

\usepackage{authblk}

\usepackage[T1]{fontenc}
\usepackage{fourier}

\usepackage{fancyhdr}
\usepackage{forloop}
\usepackage[usenames,dvipsnames]{color}
\usepackage[hyphens]{url}
\usepackage{hyphenat} 

\usepackage[pagebackref=true]{hyperref}
\definecolor{Brown}{rgb}{0.61,0.08,0.23} 
\definecolor{Blues}{rgb}{.12,.12,.48}
\hypersetup{
  colorlinks   = true,  %Colours links instead of ugly boxes
  urlcolor     = Blues, %Colour for external hyperlinks
  linkcolor    = Brown, %Colour of internal links
  citecolor    = Brown  %Colour of citations
}

\usepackage{cite}
\usepackage{multicol}
\usepackage{fancybox}
\usepackage{tikz}
\usepackage{float}

\newcommand{\gofrom}[2]{}

\newcommand{\gourl}[1]{\url{#1}}
\newcommand{\goDOI}[1]{\href{http://dx.doi.org/#1}{DOI:#1}}
\newcommand{\goScholar}[1]{{Google~Scholar}:~\href{http://scholar.google.com/scholar?cluster=#1}{#1}}

\newcommand{\partitle}[1]{\noindent {\large\textbf{\textsf{#1}}}}
\usepackage{titlesec}
\titleformat*{\section}{\Large\bf\sffamily}

\newcounter{mylabelnumber}

\newcommand{\DTM}{\ensuremath{\mbox{D-}\!\mbox{TM}}}

\newcommand{\checkendpaper}{
\setlength{\leftmargin}{0pt}
\setlength{\itemindent}{0em}

\setlength{\itemsep}{0mm}
\setlength{\parskip}{0mm}
\setlength{\parsep}{0mm} 
\linespread{0.92} 
}

\renewcommand*{\backref}[1]{}
    \renewcommand*{\backrefalt}[4]{%
    \ifcase #1 %
        (Not cited).
    \or
        (page~#2).%
    \else
        (pages~#2).
    \fi}

\usepackage{lettrine} 
\usepackage{colortbl} % Initial info-table

\makeatletter
\def\newparshape{\parshape\@npshape0{}}
\def\@npshape#1#2#3{\ifx\\#3\expandafter\@@@npshape\else\expandafter\@@npshape\fi
{#1}{#2}{#3}}
\def\@@npshape#1#2#3#4#5{%
\ifnum#3>\z@\expandafter\@firstoftwo\else\expandafter\@secondoftwo\fi
{\expandafter\@@npshape\expandafter{\the\numexpr#1+1\relax}{#2 #4 #5}{\numexpr#3-1\relax}{#4}{#5}}%
{\@npshape{#1}{#2}}}
\def\@@@npshape#1#2#3{#1 #2 }
\makeatother

\newcommand{\myquote}[2]{{
\rlap{\hspace{\dimexpr0.63\columnwidth+\columnsep}\vspace{-#1}\colorbox{black!10}{
\em\small\begin{tabular}{L{\dimexpr0.63\columnwidth-\columnsep}}
\mbox{\hspace{1pt}}\\[-1mm]
{``}#2{''}\\[2mm]
\end{tabular}}%
}}}

\title{\vspace{-20mm}\bf \Large \textsf{Software Uncertainty in Integrated Environmental Modelling: the role of Semantics and Open Science}\vspace{3mm}}
\author[1,2]{{\Large \textsf{Daniele de Rigo}} \vspace{2mm}}

\affil[1]{\small \;European Commission, Joint Research Centre, Institute for Environment and Sustainability\\

Via E. Fermi 2749, I-21027 Ispra (VA), Italy\smallskip }
\affil[2]{\;Politecnico di Milano, Dipartimento di Elettronica, Informazione e Bioingegneria\\

Via Ponzio 34/5, I-20133 Milano, Italy\vspace{-4mm}}
\date{}
\begin{document}

  \maketitle
  
\noindent\colorbox{black!10}{\small
\arrayrulecolor{white}\color{black!80}\begin{tabular}{|p{138mm}|}
\hline
\vspace{0mm}Copyright {\copyright} 2013 Daniele de Rigo.\\[6pt]
This work is licensed under a Creative Commons Attribution 3.0 Unported License \\
(\url{http://creativecommons.org/licenses/by/3.0/}).\\
See: \url{http://www.egu2013.eu/abstract_management/license_and_copyright.html}\\[6pt]
This is the author's version of the work. The definitive version has been published in the Vol. 15 of Geophysical Research Abstracts (ISSN 1607-7962) and presented at the European Geosciences Union (EGU) General Assembly 2013, Vienna, Austria, 07--12 April 2013 \\
{\url{http://www.egu2013.eu/}}\\[12pt]
Cite as:\\[7pt]
{\sloppy \parbox[l]{138mm}{
   de Rigo, D., 2013. 
   {\bf Software Uncertainty in Integrated Environmental Modelling: the role of Semantics and Open Science}. 
   {\em Geophys Res Abstr 15}, 
   \href{http://meetingorganizer.copernicus.org/EGU2013/EGU2013-13292.pdf}{13292}+ }
}\\[6pt]
Author's version DOI: \href{http://dx.doi.org/10.6084/m9.figshare.155701}{10.6084/m9.figshare.155701} , arXiv: \href{http://arxiv.org/abs/1311.4762}{1311.4762}\\[1mm]
\hline
\end{tabular}}

\vspace{9mm}\begin{mtwocols} 
\noindent \lettrine[lines=2]{C}{omputational} aspects increasingly shape environmental sciences \cite{Casagrandi_Guariso_2009}. Actually, transdisciplinary modelling of complex and uncertain environmental systems is challenging computational science (CS) and also the science-policy interface \cite{de_Rigo_F10002013,Gomes_2009,Easterbrook_Johns_2009,Hamarat_etal_2012,Bankes_2002,Kandlikar_etal_2005}.

\smallskip

Large spatial-scale problems falling within this category -- i.e. wide-scale transdisciplinary modelling for environment (WSTMe) \cite{de_Rigo_etal_EGU2013,Rodriguez_Aseretto_etal_2013,de_Rigo_etal_exp2013}
  -- often deal with factors \textbf{(a)} for which deep-uncertainty \cite{de_Rigo_F10002013,Lempert_2002,Kandlikar_etal_2005,Gober_Kirkwood_2010}
  may prevent usual statistical analysis of modelled quantities and need different ways for providing policy-making with science-based~support. 
Here, practical recommendations are proposed for tempering a peculiar -- not infrequently underestimated -- source of uncertainty. 
Software errors in complex WSTMe may subtly affect the outcomes with possible consequences even on collective environmental decision-making. 
Semantic transparency in CS \cite{de_Rigo_F10002013,de_Rigo_etal_EGU2013,de_Rigo_etal_exp2013,de_Rigo_iEMSs2012,de_Rigo_SemAP2012} and free software \cite{FSF_2012,Stallman_2009} are discussed as possible mitigations \textbf{(b)}.

\bigskip\medskip

\partitle{Software uncertainty,}\vspace{0mm}

\partitle{black-boxes and free software}

\bigskip
\noindent Integrated natural resources modelling and management (INRMM) \cite{de_Rigo_INRMM2012} frequently exploits chains of nontrivial data-transformation models ({\DTM}), each of them affected by uncertainties and errors.  

\medskip

\noindent Those {\DTM} chains may be packaged as monolithic specialized models, maybe only accessible as black-box executables (if accessible at all) \cite{Morin_etal_2012}. For end-users, black-boxes merely transform inputs in the final outputs, relying on classical peer-reviewed publications for describing the internal mechanism. 

\smallskip

\noindent While software tautologically plays a vital role in CS, it is often neglected in favour of more theoretical~\mbox{aspects}. 

\end{mtwocols}

\newcommand{\myrightdef}[1]{\ensuremath{\text{\parbox[t]{72mm}{#1}}}}
\newcommand{\myrightdefAO}[2]{\ensuremath{\text{\begin{tabular}{@{}L{#1}@{}}{\nohyphens{#2}}\end{tabular}}}}
\newcommand{\myrightdefA}[1]{\ensuremath{\hspace{3mm}\myrightdefAO{82mm}{#1}\vspace{2mm}}}
\newcommand{\myrightdefB}[1]{\ensuremath{\hspace{9mm}\myrightdefAO{76mm}{#1}\vspace{2mm}}}
\vspace{-1mm}

\begin{mtvDisplayMath}
\mbox{\textbf{(a)}} \qquad
{\begin{array}{ll}
{ \begin{array}{l}
  \mbox{Complexity} \\
  \end{array} } & { = \quad \left \{
{ \begin{array}{l}
  \myrightdefA{Transdisciplinary integration (e.g. systems of systems)}\\
  \myrightdefA{Environmental system(s) heterogeneity (e.g. geospatial fragmentation)}\\
  \myrightdefA{Data heterogeneity (formats, definitions, spatiotemporal density, ...)}\\
  \myrightdefA{Software complexity (algorithms, dependencies, languages, interfaces, ...)}\\[-2mm]
  \end{array} }
\right . } 
\\[3mm]
\\[3mm]
{ \begin{array}{l}
  \mbox{Uncertainty} \\
  \end{array} } & { = \quad \left \{
{ \begin{array}{l}
  \myrightdefA{Incomplete scientific knowledge (e.g. climate scenarios \cite{Lempert_Schlesinger_2001,Shell_2012,van_der_Sluijs_2012}, tipping points \cite{Lenton_etal_2008,Hastings_Wysham_2010,Barnosky_etal_2012},\,...\,)}\\
  \myrightdefA{Modelling assumptions and simplifications \cite{Milly_etal_2008,Sloan_Pelletier_2012,Nabuurs_etal_2008}}\\
  \myrightdefA{Uncertainty of measured/derived data} \\
  \myrightdefA{Software uncertainty} \\[-2mm]
  \end{array} }
\right . } 
\\[3mm]
\\[3mm]
{ \begin{array}{l}
  \mbox{Dynamic} \\
  \mbox{behaviour} \\
  \end{array} } & { = \quad \left \{
{ \begin{array}{l}
  \myrightdefA{Uncertainty propagation via:} \\
  \myrightdefB{Propagation in the network of interconnected WSTMe components \cite{de_Rigo_F10002013,de_Rigo_SemAP2012,Green_Sadedin_2005,de_Rigo_INRMM2012,Baker_etal_2012,de_Rigo_etal_IPRMW2012,Thompson_etal_2009,FAO_2005,Bonan_2008}} \\
  \myrightdefB{Iterations within nonlinear optimization steps \cite{Hamarat_etal_2012,Ferreira_etal_2012,de_Rigo_2001,Bond_etal_2011,de_Rigo_etal_2005,Phillis_Kouikoglou_2012,Cavallo_Nardo_2008,Castelletti_etal_2008}} \\
  \myrightdefB{Data fusion, harmonization, integration \cite{Rodriguez_Aseretto_etal_2013,Kempeneers_etal_2011,Sedano_etal_2012,de_Rigo_Bosco_2011,Voinov_Shugart_2013}} \\
  \myrightdefB{Steps for computing and aggregating criteria and indices \cite{Bankes_2002,Lempert_2002,Kandlikar_etal_2005,Mendoza_Martins_2006,OFarrell_Anderson_2010,Dale_Beyeler_2001,Gilbert_2010}} \\[-2mm]
  \end{array} }
\right . } 
\end{array}}
\end{mtvDisplayMath}
\vspace{0mm}

\begin{mtwocols} 
\noindent This paradox has been provocatively described as ``the invisibility of software in published science. Almost all published papers required some coding, but almost none mention software, let alone include or link to source code'' \cite{Barnes_Jones_2011}.
\end{mtwocols} 

\begin{mtwocols} 
\noindent Recently, this primacy of theory over reality  \cite{Sanders_Kelly_2008,Cerf_2012,Pincas_2011} has been challenged by new emerging hybrid approaches \cite{Sanders_2009} and by the growing debate on open science and scientific knowledge freedom  \cite{de_Rigo_F10002013,Kleiner_2011,Nature_2011,Peng_2011,Cai_etal_2012}. 

In particular, the role of free software has been underlined within the paradigm of reproducible research \cite{Morin_etal_2012,Peng_2011,Cai_etal_2012,Ghisla_etal_2012}. 
In the spectrum of reproducibility, the free availability of the source code is emphasized \cite{Peng_2011} as the first step from non-reproducible research (only based on classic peer-reviewed publications) toward reproducibility.
\end{mtwocols}

\newcommand{\mtvbox}[1]{\ensuremath{
\text{\colorbox{red!15}{$#1$}}
}}
\newcommand{\mtvsem}[2]{::\!\!\left | \,{#1}\, \right |\!\!::^{\begin{array}{l}
\,^{\kern-1em #2}\\
\end{array}}}
\newcommand{\ought}{\Box}

\begin{mtvDisplayMath}
{ \begin{array}{l}
\quad{ \begin{array}{ll}
Y \quad=\quad f^*(\,X\,) \quad=\quad f(\, \theta^*\,,\, X \,)    &
\myrightdef{
Theoretic {\DTM} whose algorithm is typically described in peer reviewed publications. The {\DTM} may e.g. implement a given WSTMe as instance of a suitable family of functions $f$ by means of selected parameters $\theta^*$. $\theta^*$ may be the result of an optimization (regression, control problem, ...).
}\\[3mm] 
\\[0mm]
Y \quad=\quad f^{\mtvbox{\zeta}} \quad=\quad f(\, \theta^{\mtvbox{\zeta}}\,,\, X\,,\, \mtvbox{\zeta}\,\, ) \hspace{6mm}&
\myrightdef{
Real {\DTM} where the software uncertainty $\mtvbox{\zeta}$ may affect both the function family $f$ and the optimality of the selected parameters $\theta^\zeta$.
}\\[3mm] 
\\[0mm]
\mtvsem{f(\,\theta\,,\,X\,,\,\mtvbox{\zeta}\,\,)}{sem}&
\myrightdef{
Semantically enhanced {\DTM} (e.g. SemAP). The {\DTM} is subject to the semantic checks $sem$ as pre-, post-conditions and invariants on inputs,  outputs and the {\DTM} itself:
}\\[3mm] 
\\[0mm]
&{ Y = \;\;\mtvsem{f(\theta,X,\zeta\,)}{sem} \Leftrightarrow \left \{ { \begin{array}{l}
Y=f(\theta,X,\zeta\,) \\[1mm]
\ought sem( Y, f, \theta,X,\zeta\, ) \\
\end{array}} \right . }
\end{array} }\\
\mbox{\textbf{(b)}}
\\[5mm]
\quad\quad\mbox{where } { \left \{ \begin{array}{l}
 X \mbox{ is the input array of data }
 X = \{ X_{1}, X_{2}, \cdots X_{i} \cdots X_{n} \}\\[2mm]
 X_{i} \in \mathbf{C}^{N_{i1} \times \cdots \times N_{in_i}} 
\mbox{ is a multi-dimensional array (e.g. a two-dimensional } \\
  \mbox{\qquad raster layer)} \\[2mm]
 Y \mbox{ is analogously the output array of data } \\[2mm]
\mbox{the modal/deontic logic operator \;} {\ought p } \mbox{\; means: it ought to be that } p .
\end{array} \right . }
\end{array} }
\end{mtvDisplayMath}

\begin{mtwocols}
\noindent Applying this paradigm to WSTMe, an alternative strategy to black-boxes would suggest exposing not only final outputs but also key intermediate layers of data and information along with the corresponding free software {\DTM} modules. 
\end{mtwocols}

\begin{mtwocols}  
\newparshape{12}{0pt}{0.63\columnwidth}{1}{0pt}{\columnwidth}\\
\myquote{50mm}{Software errors in complex WSTMe may subtly affect the outcomes with possible consequences even on collective environmental decision-making''\\[3.2mm]
\hline \\[0.2mm]
``The chain of free-software modules should be transparent}
\noindent A concise, semantically-enhanced modularization \cite{de_Rigo_iEMSs2012,de_Rigo_SemAP2012} may help not only to see the code (as a very basic prerequisite for semantic transparency) but also to understand -- and correct -- it \cite{Iverson_1980}. Semantically-enhanced, concise modularization is e.g. supported by semantic array programming (SemAP) \cite{de_Rigo_iEMSs2012,de_Rigo_SemAP2012} and its extension to geospatial problems \cite{de_Rigo_etal_EGU2013,de_Rigo_etal_exp2013}. 

\smallskip

\newparshape{6}{0pt}{\columnwidth}{12}{0.37\columnwidth}{0.63\columnwidth}{1}{0pt}{\columnwidth}\\
Some WSTMe may surely be classified in the subset of software systems which ``are growing well past the ability of a small group of people to completely understand the content'', while ``data from these systems are often used for critical decision \mbox{making}''~\cite{Sanders_Kelly_2008}.  
In this context, the further uncertainty arising from the unpredicted ``(not to say unpredictable)'' \cite{Cerf_2012} behaviour of software errors propagation in WSTMe should be explicitly considered as software uncertainty~\cite{Lehman_1989,Lehman_Ramil_2002} (see {\bf b}). 
The data and information flow of a black-box {\DTM} is often a (hidden) composition of {\DTM} modules:
\begin{figure}[H]
\vspace{-0.5mm} 
\centerline{\includegraphics[scale=0.18]{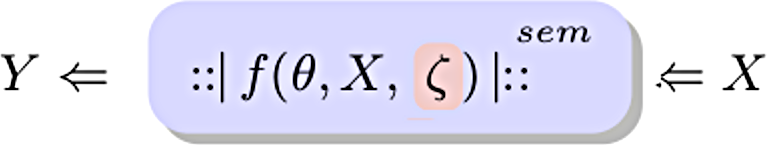}}
\vspace{-2.5mm} 
\end{figure}
\noindent This chain of free-software {\DTM} modules
(each of them semantically-enhanced) should be transparent:
\end{mtwocols}

\begin{figure}[H]
\vspace{6mm} 
\centerline{\includegraphics[scale=0.18]{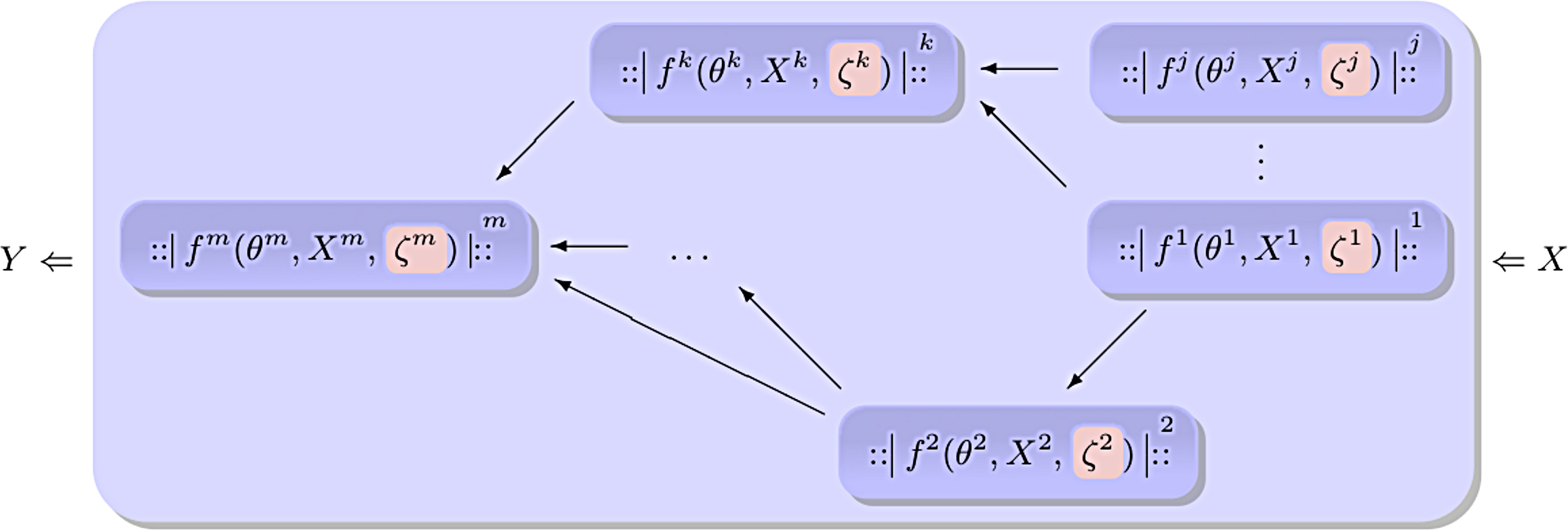}}
\vspace{5mm} 
\end{figure}
\bigskip

\begin{mtwocols}
\partitle{Semantics and design diversity}
\bigskip\vspace{0mm}

\noindent Silent faults \cite{Hook_Kelly_2009} are a critical class of software errors altering computation output without evident symptoms -- such as computation premature interruption (exceptions, error messages, ...), obviously unrealistic results or computation patterns (e.g. noticeably shorter/longer or endless computations). As it has been underlined, ``many scientific results are corrupted, perhaps fatally so, by undiscovered mistakes in the software used to calculate and present those results''~\cite{Hatton_2007}. 

\newparshape{15}{0pt}{0.63\columnwidth}{4}{0pt}{\columnwidth}{15}{0.37\columnwidth}{0.63\columnwidth}{1}{0pt}{\columnwidth}\\
\myquote{67mm}{\vspace{0mm}Semantic modularization might help to catch at least a subset of silent faults, when misusing intermediate data outside the expected semantic context''\\[4mm]
\hline \\[1mm]
``Where the complexity and scale may lead to deep uncertainty, techniques such as ensemble modelling may be recommendable\vspace{1mm}}
\noindent Despite the ubiquity of software errors~\cite{Lehman_1989,Lehman_Ramil_2002,Hook_Kelly_2009,Hatton_2007,Hatton_1997,Hatton_2012,Lehman_1996,Oberkampf_etal_2002,Wilson_2006}, the structural role of scientific software uncertainty seems dramatically underestimated  \cite{de_Rigo_F10002013,Cerf_2012}. Semantic {\DTM} modularization might help to catch at least a subset of silent faults, when misusing intermediate data outside the expected semantic context of a given {\DTM} module \textbf{(b)}. 
Where the complexity and scale of WSTMe may lead unavoidable software-uncertainty to induce or worsen deep-uncertainty \cite{de_Rigo_F10002013}, techniques such as ensemble modelling may be recommendable \cite{Lempert_2002,Kandlikar_etal_2005,Gober_Kirkwood_2010}. Adapting those techniques for glancing at the software-uncertainty of a given WSTMe would imply availability of multiple instances (implementations) of the same abstract WSTMe. 
Independently re-implementing the same WSTMe (design diversity \cite{Rebaudengo_2011}) might of course be extremely expensive. However, partly independent re-implementations of critical {\DTM} modules may be more affordable and examples of comparison between supposedly equivalent {\DTM} algorithms seem to corroborate the interest of this research option \cite{Cai_etal_2012,Beaudette_2008,Barnes_Jones_2011}.
\end{mtwocols}

\medskip
\checkendpaper{}

\begin{footnotesize}

\renewcommand*\labelenumi{[\theenumi]}

\raggedright
\nohyphens{

}
\end{footnotesize}

\end{document}